\documentclass[doublecol]{epl2}
\usepackage[a4paper,top=3cm,bottom=2.6cm,left=1.2cm,right=1.8cm]{geometry}

\usepackage{amsmath}
\usepackage{amssymb}
\usepackage{bigfoot}
\usepackage{booktabs}
\usepackage{graphicx}
\usepackage[bookmarks=false]{hyperref}
\usepackage[utf8]{inputenc}
\usepackage{newunicodechar}

\renewcommand{\year}{0}
\usepackage{pgf}

\renewcommand{\vec}[1]{\boldsymbol{\mathbf{#1}}}

\newunicodechar{−}{\ensuremath{-}}

\hypersetup{
  pdfstartview = FitH,
  pdftoolbar   = false,
  pdfmenubar   = true,
  colorlinks   = true,
  linkcolor    = blue!50!black,
  urlcolor     = blue!50!black,
  citecolor    = blue!50!black
}

\title{{\AA}ngstr{\"o}m-scale chemically powered motors}
\author{Peter H. Colberg \and Raymond Kapral}
\shortauthor{Peter H. Colberg and Raymond Kapral}
\institute{
  Chemical Physics Theory Group, Department of Chemistry,
  University of Toronto\\
  Toronto, Ontario M5S 3H6, Canada
}

\abstract{
Like their larger micron-scale counterparts, {\AA}ngstr{\"o}m-scale chemically
self-propelled motors use asymmetric catalytic activity to produce
self-generated concentration gradients that lead to directed motion.
Unlike their micron-scale counterparts, the sizes of {\AA}ngstr{\"o}m-scale
motors are comparable to the solvent molecules in which they move, they are
dominated by fluctuations, and they operate on very different time scales.
These new features are studied using molecular dynamics simulations of small
sphere dimer motors.
We show that the ballistic regime is dominated by the thermal speed but the
diffusion coefficients of these motors are orders of magnitude larger than
inactive dimers.
Such small motors may find applications in nano-confined systems or perhaps
eventually in the cell.
}

\begin{document}

\maketitle

Biological molecular motors operate under non-equilibrium conditions and consume
fuel in their environment to drive conformational changes that enable them to
carry out specific tasks.
Motor proteins such as adenosine triphosphate synthase pump ions
across the cell membrane, kinesin walks along microtubules to effect active
transport in the cell, to name just two of the thousands of motors that
contribute to the biological function of the cell~\cite{alberts-cell}.
These small molecular motors are able to perform their tasks in complex
environments in spite of strong thermal fluctuations.
Synthetic mimics of such molecular machines have been
constructed~\cite{pei:06,yin:08,omabegho:09}.
On much larger micron scales, bacteria swim by using chemical fuel to drive
various types of non-reciprocal conformational changes to produce directed
motion~\cite{purcell:77}.

Chemically powered synthetic motors that operate by phoretic mechanisms and do
not rely on conformational changes for directed motion have been the subject of
recent studies~\cite{hong:10,mirkovic:10a,wang:09b,kapral:13}.
Metallic rod~\cite{Paxton2004, Zacharia2009}, Janus
particle~\cite{Ebbens2012,Baraban2012} and sphere dimer~\cite{Valadares2010}
motors with linear dimensions in the hundreds of nanometres and micron ranges
have been extensively investigated and hold the promise of future applications,
which include targeted cargo transport, controlled motion and stirring in
microfluidic arrays, active self-assembly and directed chemical synthesis.
Although such small motors are influenced by thermal fluctuations, continuum
descriptions of phoretic motion~\cite{Anderson1984,Anderson1991,Fair1989}
adequately describe the dynamics of such
motors~\cite{golestanian:05,julicher:09,popescu:10,sabass:11,Ebbens2012}.

In this article, we consider the dynamics of even smaller {\AA}ngstr{\"o}m-scale
chemically powered motors with sizes of a few nanometres where new
factors come into play.
The motors are no longer much larger than the solvent molecules that comprise
their environments.
Consequently, solvent structure in the vicinity of the motor must be taken into
account.
Motors are subject to orientational fluctuations that tend to destroy ballistic
motion.
Since most potential applications utilize ballistic motion, the long
reorientational times of mesoscale motors make directed motion easy to observe.
For small {\AA}ngstr{\"o}m-scale motors the reorientational times are very
short.
Straightforward application of macroscopic and hydrodynamic models to the
dynamics is questionable or, at the very least, must be examined carefully.
Nevertheless, it is well known that collective hydrodynamic modes play a crucial
part in molecular-level dynamics, an example being the long-time tail in the
velocity autocorrelation function~\cite{alder:67}.
Finally, fluctuations play an even more dominant role for these small motors
than for larger mesoscale motors.
Thus, the investigation of the dynamics of molecular-scale motors is
both interesting and presents fundamental challenges.

Additional stimulus for such research is provided by potential applications of
synthetic motors inside the cell or in very small nanoscale environments.
Recently, experiments have shown that single enzymes and small
organometallic molecules undergoing catalytic activity exhibit enhanced
diffusion in comparison to their inactive forms, and this enhancement has been
attributed to propulsion on the molecular level~\cite{muddana:2010,
sengupta:2013, pavlick:2013}.
Given this context, studies of the dynamics of small molecular-scale
motors will contribute to the origins, characterization and consequences of
directed motion in this regime.

\begin{figure}[t]
  \includegraphics[width=\linewidth]{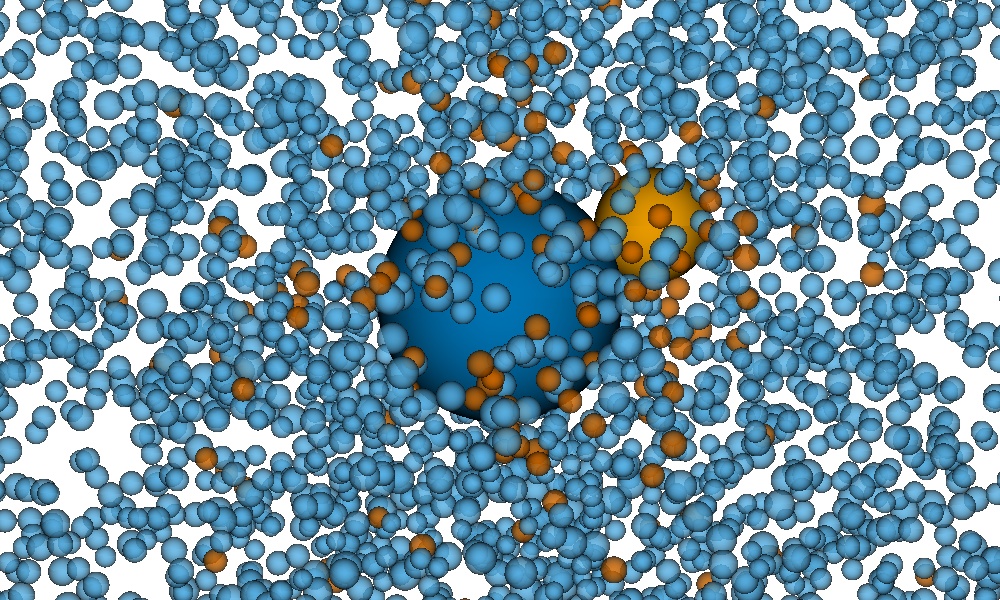}
  \caption{Sphere dimer motor ($\sigma_\text{N} = 10$), comprising
  C~sphere (orange) and N~sphere (blue), in solvent consisting of
  species A (sky blue) and B (vermilion).
  The solvent is shown with reduced density $0.04\sigma^{-3}$ instead of
  $0.8\sigma^{-3}$.}
  \label{fig:dimer}
\end{figure}

We employ full molecular dynamics to study a simple model system made of a small
sphere dimer motor~\cite{Rueckner2007}, comprising linked catalytic and
non-catalytic spheres, in a structureless solvent (see fig.~\ref{fig:dimer}).
Chemical reactions at the catalytic sphere convert reactants to products.
Due to the spatially asymmetric catalytic activity of the motor, an
inhomogeneous distribution of these species is produced.
This self-generated concentration gradient is an essential element that leads
to propulsion in solution~\cite{kapral:13}.
Micron-size sphere dimer motors comprising silica non-catalytic and Pt catalytic
spheres have been studied experimentally~\cite{Valadares2010}.
While motor dynamics often takes place in aqueous solution, our structureless
solvent model is sufficient to capture the important dynamical and structural
effects that occur at the {\AA}ngstr{\"o}m scale.
Applications to specific systems should address effects related to the unique
structural properties of water.

The system is contained in a cube of volume $V = (50\,\sigma)^3$ with
periodic boundary conditions.
The solvent consists of $N_\text{s} = 10^5$ particles with
number density $\varrho_\text{s} = N_\text{s} / V =
0.8\sigma^{-3}$ and two species,
reactant (A) and product (B), both with diameter $\sigma$ and mass $m$.
The motor consists of rigidly linked catalytic (C) and non-catalytic
(N) spheres with diameters $\sigma_\text{C} = 2\sigma, \dots, 5\sigma$ and
$\sigma_\text{N} = 4\sigma, \dots, 10\sigma$, respectively,
and neutrally buoyant mass $M_\text{m} = \frac{\pi}{6} \varrho_\text{s} m
(\sigma_\text{C}^3 + \sigma_\text{N}^3)$.
The separation of the spheres is chosen as\footnote{
  The choice of $R$ conserves energy for the chemical reaction
  $\text{A}\rightarrow\text{B}$ in light of the differing $\epsilon_\text{NA}$
  and $\epsilon_\text{NB}$, by ensuring that a solvent particle A or B is within
  interaction range of at most a single dimer sphere, either C or N, at any
  given time.
}
$R = \sqrt[6]{2}\left(\left(\sigma_\text{C} + \sigma_\text{N}\right)/2 +
\sigma\right)$.
The C sphere catalyses the chemical reaction\footnote{
  The reaction is diffusion-limited since $k = k_D k_0 / (k_D + k_0) \simeq
  k_D$, where the reaction rate constant $k_0 = R_0^2\sqrt{8\pi k_\text{B}T/m}$
  is 8--16 times larger than the Smoluchowski rate constant $k_D = 4\pi
  D_\text{s} R_0$ for $R_0 = 2^{1/6} (\sigma_\text{C} + \sigma) / 2$ and
  $D_\text{s} = 0.086$.
}
$\text{A}\rightarrow \text{B}$, which occurs with unit probability when A
comes within interaction range of C.
All particles interact via the shifted, truncated Lennard-Jones potential,
$V_{ij}(r) = \epsilon_{ij}(4((\sigma_{ij}/r)^{12} -(\sigma_{ij}/r)^6) + 1)$
for $r < \sqrt[6]{2}\,\sigma_{ij}$ and zero otherwise.
Here $r$ is the minimum image distance between the centres of a pair of
particles, $\sigma_\text{CA} = \sigma_\text{CB} =
\frac{1}{2}\left(\sigma_\text{C}+\sigma\right)$ and $\sigma_\text{NA} =
\sigma_\text{NB} = \frac{1}{2}\left(\sigma_\text{N}+\sigma\right)$ for pairs
of dimer sphere and solvent particle, and $\sigma_\text{AA} =
\sigma_\text{AB} = \sigma_\text{BB} = \sigma$ for pairs of solvent particles.
The interaction energy is $\epsilon$ for all pairs apart from NB pairs,
where $\epsilon_\text{NB} = 0.1\epsilon, \dots, 10\epsilon$.
The temperature of the system is $k_\text{B}T/\epsilon = 1$.
The system is maintained in a non-equilibrium steady state by converting B
particles back to A far from the dimer.
Newton's equations of motion were solved\footnote{
  The simulations were performed on GPUs and CPUs using a self-written
  code in the programming languages OpenCL~C and Lua to be
  published alongside a subsequent computational article.
}
iteratively using the velocity-Verlet algorithm with $\delta t = 0.001\,\tau$.

Simulation results are reported in dimensionless units where distance is given
in units of $\sigma$, mass in units of $m$, energy in units of $\epsilon$, and
time in units of $\tau = \sigma\sqrt{m\epsilon^{-1}}$.
Our parameters are chosen to model a dense fluid argon-like solvent and, using
argon~\cite{Rahman1964} values of $\sigma = 0.34\,\text{nm}$, $\epsilon =
120\,\text{K}\,k_\text{B}$, and $m = 39.95\,\text{u}$ and $\tau =
2.15\,\text{ps}$, we can assign physical values to our {\AA}ngstr{\"o}m-scale
motor simulations.

\begin{figure}[htbp]
  \input{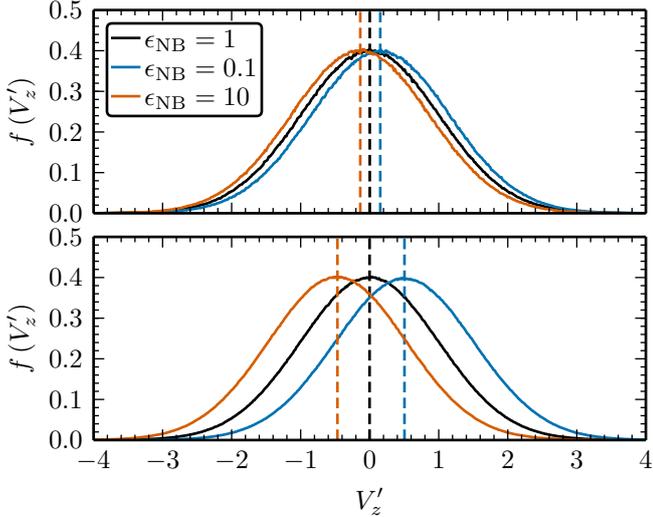}
  \caption{Probability densities of the scaled propulsion velocity, $f(V_z')$,
  for small ($\sigma_\text{N} = 4$; top) and large ($\sigma_\text{N} = 10$;
  bottom) dimers, for several values of $\epsilon_\text{NB}$.
  The dashed lines indicate the means, $\langle V_z' \rangle$, of the
  distributions.}
  \label{fig:velocity}
\end{figure}

The non-equilibrium steady-state average velocity of the sphere dimer motor
projected along its internuclear axis, $\langle V_z \rangle$, can be used to
characterize the dynamics of the motor.
Here $V_z=\hat{{\bf z}} \cdot {\bf v}_\text{cm}$, with ${\bf v}_{cm}$ the
velocity of the centre of mass of the dimer and $\hat{{\bf z}}$ a unit vector
along the bond ($z$-axis) directed from the N sphere to the C sphere.
The average $\langle V_z\rangle$ and its fluctuations can be determined from
the probability density, $f(V_z)$, which is shown in fig.~\ref{fig:velocity} for
two dimers with different sizes, $\sigma_\text{N} = 10$ and $\sigma_\text{N} =
4$, for several values of $\epsilon_\text{NB}$.
The figure plots the scaled velocity $V_z'=V_z\sqrt{M_\text{m}/k_\text{B}T}$ so
that in scaled units the widths of the distributions are the same.
These probability densities have a Maxwell-Boltzmann form,
$f(V_z')=(2 \pi)^{-1/2}{\mathrm e}^{-(V_z'-\langle V_z' \rangle)^2/2}$.
The figure shows that the sphere dimer moves in the direction of
the catalytic sphere (positive $\langle V_z' \rangle$) when $\epsilon_\text{NB} < 1$
and in the opposite direction (negative $\langle V_z' \rangle$) when
$\epsilon_\text{NB} > 1$; when $\epsilon_\text{NB}=1$, $\langle V_z \rangle =0$.
The values of $\langle V_z' \rangle$ are given in Table~\ref{tab:dimer} for a
variety of dimer sizes and potential parameters.
As the dimer size decreases the propulsion speed is a smaller fraction of the
mean thermal speed and the effects of fluctuations are felt more strongly.
\begin{table}[htbp]
  \centering
  \setlength{\tabcolsep}{5pt}
  \begin{tabular}{ccrrrrrr}
    \toprule
    $\sigma_\text{N}$
    & $\epsilon_\text{NB}$
    & \multicolumn{1}{c}{$\langle V_z^\prime\rangle$}
    & \multicolumn{1}{c}{$\tau_\text{r}$}
    & \multicolumn{1}{c}{$D_\text{m}$}
    & \multicolumn{1}{c}{$D_\text{m}^\text{th}$}
    & \multicolumn{1}{c}{${\mathcal B}_\text{I}^\prime$}
    & \multicolumn{1}{c}{${\mathcal B}_\text{I}^{\prime\text{th}}$}
    \\
    \midrule
    10 & 1 & -0.002 & 2035 & 0.006 &  & 2.977 &  \\
    10 & 0.1 & 0.504 & 1931 & 0.361 & 0.352 & 3.265 & 3.254 \\
    10 & 10 & -0.468 & 3774 & 0.659 & 0.590 & 3.196 & 3.219 \\
    \midrule
    8 & 1 & 0.001 & 1010 & 0.008 &  & 2.985 &  \\
    8 & 0.1 & 0.365 & 1078 & 0.198 & 0.207 & 3.138 & 3.133 \\
    8 & 10 & -0.345 & 2134 & 0.389 & 0.359 & 3.098 & 3.119 \\
    \midrule
    6 & 1 & 0.002 & 531 & 0.010 &  & 2.981 &  \\
    6 & 0.1 & 0.254 & 503 & 0.109 & 0.117 & 3.062 & 3.065 \\
    6 & 10 & -0.232 & 1036 & 0.207 & 0.192 & 3.025 & 3.054 \\
    \midrule
    4 & 1 & 0.001 & 195 & 0.018 &  & 2.952 &  \\
    4 & 0.1 & 0.153 & 174 & 0.057 & 0.063 & 2.991 & 3.023 \\
    4 & 10 & -0.137 & 302 & 0.110 & 0.080 & 2.963 & 3.019 \\
    \bottomrule
  \end{tabular}
  \caption{Sphere dimer properties for several diameters, $\sigma_\text{N}$
  and $\sigma_\text{C} = \sigma_\text{N} / 2$, and values of
  $\epsilon_\text{NB}$:
  Mean scaled propulsion velocity, $\langle V'_z\rangle$, orientational
  relaxation time, $\tau_\text{r}$, diffusion constant, $D_\text{m}$,
  theoretical estimate for diffusion constant, $D_\text{m}^\text{th}$,
  scaled ballistic prefactor, ${\mathcal B}_\text{I}^\prime$, and theoretical estimate
  for scaled ballistic prefactor, ${\mathcal B}_\text{I}^{\prime\text{th}}$.}
  \label{tab:dimer}
\end{table}

The propulsion of the sphere dimer has its origin in the force on the dimer that
arises from the different intermolecular potentials of the A and B particles
with the dimer spheres and the self-generated non-equilibrium inhomogeneous
steady-state concentration fields.
As a result the non-equilibrium average of the force projected along the dimer
axis is $\langle \hat{{\bf z}} \cdot {\bf F}_\text{m} \rangle \ne 0$.
Due to momentum conservation, this force can be written in terms of the force
exerted on the solvent,
\begin{alignat}{1}
  \label{t2force}
  \langle \hat{{\bf z}} \cdot {\bf F}_\text{m} \rangle =&
  \int {\mathrm d}{\bf r}~\varrho(\mathbf{r})~(\mbox{$\hat{\mathbf{z}}$} \cdot \hat{\mathbf{r}})
  \frac{dV_\text{CA}(r)}{dr} \nonumber
  \\
  &
  + \sum_{\alpha=A}^{B}~\int {\mathrm d}{\bf r}'~\varrho_\alpha(\mathbf{r}'+ R \hat{\mathbf{z}})~(\mbox{$\hat{\mathbf{z}}$}
  \cdot \hat{\mathbf{r}}')
  \frac{dV_{\text{N}\alpha}(\mathbf{r}')}{dr'}\,,
\end{alignat}
where $\mathbf{r}$ has the catalytic C sphere as the origin, while $\mathbf{r}'
= \mathbf{r} - R \mbox{$\hat{\mathbf{z}}$}$ is defined with the non-catalytic N
sphere as the origin.
In eq.~(\ref{t2force}) $\varrho_\alpha(\mathbf{r})=\langle \varrho_\alpha(\mathbf{r};
{\bf r}^{N_\alpha}) \rangle $ is the non-equilibrium average of the microscopic
concentration field, $\varrho_\alpha(\mathbf{r}; {\bf
r}^{N_\alpha})=\sum_{i=1}^{N_\alpha}\delta({\bf r}_{i \alpha}-{\bf r})$.
The total concentration of the A and B fields is $\varrho({\bf r})
= \varrho_\text{A}({\bf r}) + \varrho_\text{B}({\bf r})$.
In our simulations we have $V_\text{CA}=V_\text{CB}$ and have used this fact in
writing the first term in eq.~(\ref{t2force}), which now depends only on the
total concentration in the vicinity of the C sphere.
Since $V_\text{NA} \ne V_\text{NB}$, the second term depends on the individual A
and B concentration fields close to the N sphere and is the term that dominates
the force.

\begin{figure}[htbp]
  \input{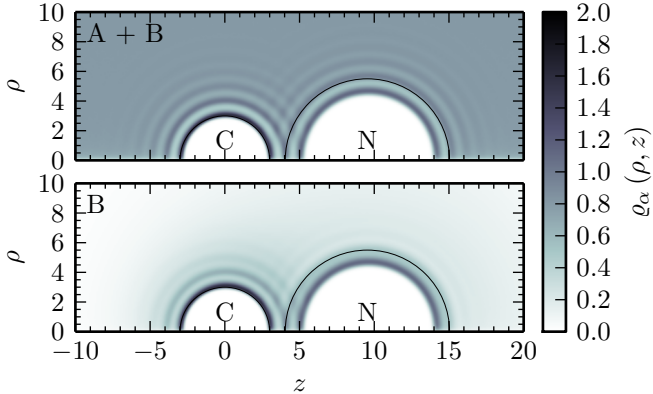}
  \caption{Cylindrically averaged solvent density fields, $\varrho({\bf r})$
  (top) and $\varrho_\text{B}({\bf r})$ (bottom), in the frame of the moving dimer,
  for the large dimer ($\sigma_\text{N} = 10$) and $\epsilon_\text{NB} = 0.1$.
  The solid black lines indicate the interaction distances
  $\sigma_{\text{C}\alpha}$ and $\sigma_{\text{N}\alpha}$.}
  \label{fig:cylindrical_density}
\end{figure}
\begin{figure}[htbp]
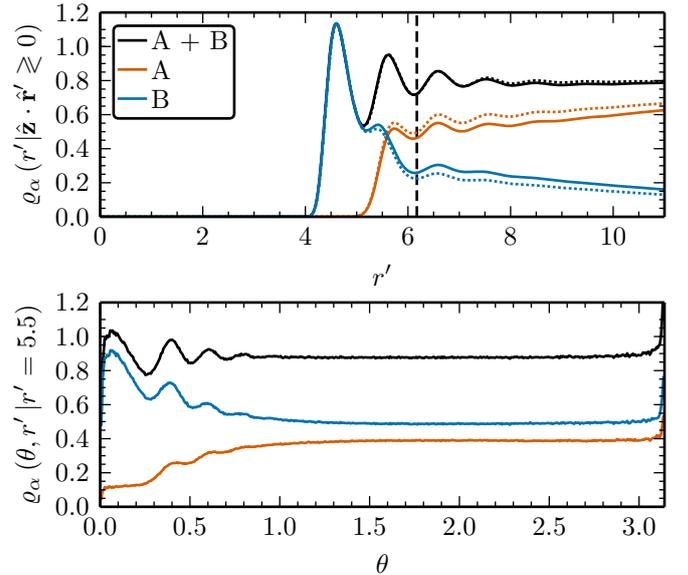

  \input{hemispherical_density}
  \input{polar_density}
  \caption{Radial (top) and angular (bottom) solvent density, of species A, B, and
  both, from the centre of the N sphere, for the large dimer ($\sigma_\text{N} =
  10$) and $\epsilon_\text{NB} = 0.1$.
  The radial density is averaged separately over each hemisphere: the solid lines
  show the density averaged over the hemisphere on the positive $z$-axis
  (towards the C sphere), the dashed lines over the hemisphere on the negative
  $z$-axis (away from the C sphere).
  The vertical dashed line indicates the cut-off distance of the force,
  $r_\text{c} = \sqrt[6]{2}\left(\sigma_\text{N} + 1\right) / 2$.
  The angular density is averaged over the azimuth angle, $\phi$, at a distance
  $r' = 5.5$ from the centre of the N sphere, which corresponds to the second
  peak of the total radial density.}
  \label{fig:radial_density}
\end{figure}

The concentration fields are significantly influenced by solvent structural
effects as can be seen in fig.~\ref{fig:cylindrical_density}, which shows
$\varrho({\bf r})$ and $\varrho_\text{B}({\bf r})$ in a cylindrical coordinate
frame $(\rho,z,\phi)$ that is co-moving with the dimer.
The concentration fields exhibit oscillations near the surface of the dimer
spheres that are characteristic of particles with an excluded volume.
The radial (top) and angular (bottom) concentration fields with
origin at the centre of the N sphere are shown in fig.~\ref{fig:radial_density}.
The radial concentration fields are determined by separately averaging over N
hemispherical regions with $\hat{{\bf r}}'\cdot \hat{{\bf z}} > 0$ (towards the C
sphere, solid lines) and $\hat{{\bf r}}'\cdot \hat{{\bf z}} < 0$ (away from the
C sphere, dashed lines) in order to characterize the inhomogeneous nature
of the concentration fields around the N sphere.
Since the interaction potentials have finite range and are non-zero only for
$r' < \sqrt[6]{2} \sigma_{\text{N}\alpha}$ (vertical dashed line in the figure),
the density fields with values in this range contribute to the propulsion force;
the system is force-free outside this boundary layer region.
The pronounced oscillations in the radial concentration fields are evident
within the boundary layer surrounding the dimer and directly affect the
propulsion force in eq.~(\ref{t2force}).
A major element in the force that drives propulsion is the concentration
inhomogeneity across the N sphere and this is evident in the angular dependence
of the concentration fields at a fixed value of $r'$ shown in the lower panel of
the figure.
For $\theta$ values close to zero near the C sphere the B concentration is large
and the A concentration is small.
As $\theta$ increases to values near $\pi$ on the opposite side of the N
sphere the A concentration increases while B decreases.

\begin{figure}[htbp]
  \input{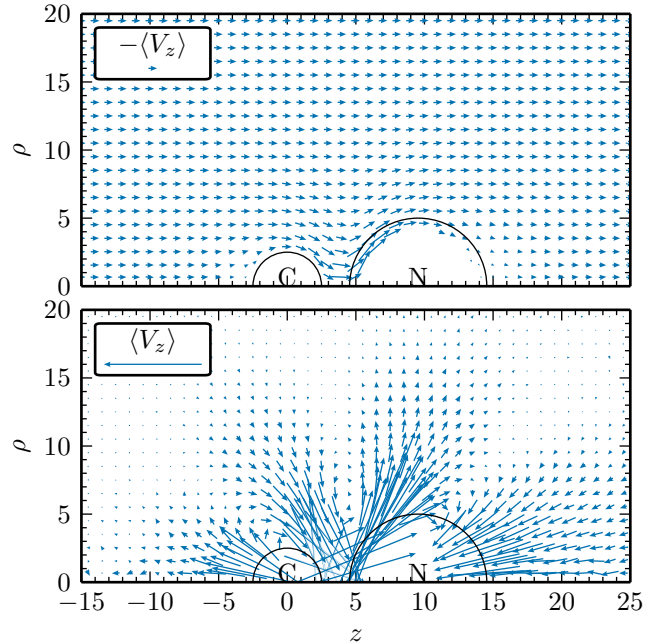}
  \caption{Cylindrically averaged fluid velocity field, $\vec{v}\left(\rho, z\right)$,
  relative to the moving dimer, for the large dimer ($\sigma_\text{N} = 10$) and
  $\epsilon_\text{NB} = 0.1$.
  The fluid velocity field is shown in the dimer frame (top) and in the
  simulation frame (bottom).}
  \label{fig:cylindrical_velocity}
\end{figure}

Due to momentum conservation, fluid flows are generated in
the solvent and are an integral part of the propulsion mechanism.
On molecular scales continuum hydrodynamic approaches break down.
However, such descriptions have proven to be successful on even very small
scales.
Similar observations apply to the flows generated by our {\AA}ngstr{\"o}m-scale
motors.
Due to the small size of a motor and the strong thermal fluctuations it
experiences, extensive averaging is required to visualize the fluid flow
fields\footnote{
  The velocity fields are averaged over 1000 configurations per
  realization of $10^7$ steps, and 100 realizations.
  The solvent velocities are transformed to the non-inertial dimer frame
  by $\vec{v}'_i = \vec{v}_i - \vec{v}_\text{cm} -
  \vec{\omega}\times\left(\vec{r}_i - \vec{r}_\text{cm}\right)$,
  where $\omega$ is the angular velocity of the dimer.
}.
In fig.~\ref{fig:cylindrical_velocity} the fluid velocity fields are shown in a
cylindrical coordinate fame using two different representations.
The top panel shows the flow field in a moving frame where the dimer velocity is
zero.
The flow field far from the dimer has the value $-\langle V_z \rangle$.
The lower panel shows the flow field in the vicinity of the moving dimer.
Here the flow field far from the dimer tends to zero while a flow field with
dipole-like components exists in the vicinity of the dimer.
These results are consistent with the macroscopic flows that appear in the
phoretic mechanism for propulsion~\cite{anderson:83}.

\begin{figure}[htbp]
  \input{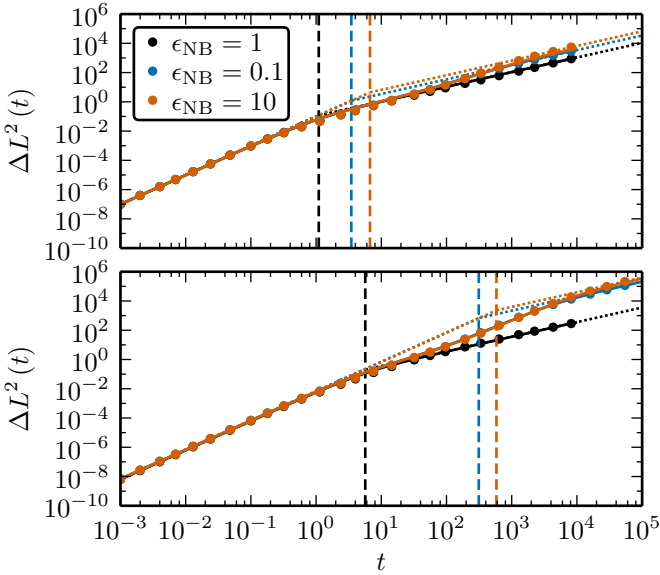}
  \caption{Mean square displacement for small ($\sigma_\text{N} = 4$; top) and
  large ($\sigma_\text{N} = 10$; bottom) dimers, for several values of
  $\epsilon_\text{NB}$.
  The dotted lines show fits of ballistic and diffusive regimes.
  The vertical dashed lines indicate crossover times, $\tau_\text{c}$, between
  these regimes.
  The solid lines show theoretical estimates (eq.~(\ref{eq:MSD-th})).}
  \label{fig:msd}
\end{figure}

The mean square displacement (MSD) of the motor, $\Delta L^2(t) =
\langle |{\bf r}_\text{cm}(t)-{\bf r}_\text{cm}(0)|^2 \rangle$, where
${\bf r}_\text{cm}$ is the position of the centre of mass of the dimer, is
plotted in fig.~\ref{fig:msd} for chemically active dimers with
different sizes, $\sigma_\text{N}=10$ and $\sigma_\text{N}=4$; the results for the
corresponding chemically inactive dimers are also plotted for comparison.
Both ballistic ($\sim {\mathcal B}_\text{I} t^2$) and diffusive ($\sim 6 D_\text{m} t$)
regimes, indicated by the straight lines, can be seen in the plots.
The vertical dashed lines indicate the crossover times, $\tau_\text{c}$, between these
regimes.
The prefactors ${\mathcal B}'_\text{I}={\mathcal B}_\text{I} M_\text{m}/k_\text{B} T$ that
characterize the ballistic regime and the dimer diffusion coefficients $D_\text{m}$,
obtained from short- and long-time fits to the data, respectively, are given in
Table.~\ref{tab:dimer}, along with several other quantities that are used in the
analysis given below.

The MSD can be written using the dimer velocity autocorrelation function (VAF),
$C_{VV}\left(t\right)
  = \frac{1}{3}\langle\vec{V}\left(t\right)\cdot\vec{V}\rangle$, as
\begin{equation}
  \Delta L^2(t)
  = \int_0^{t} {\mathrm d}t^{\prime}\, \int_0^{t^\prime}
  {\mathrm d}t^{\prime \prime}\,C_{VV}\left(t^{\prime \prime}\right).
\end{equation}
Expressing the velocity in terms of its average in the direction of the dimer
axis and deviations from this value, $\vec{V}\left(t\right) = \langle V_z\rangle
\hat{{\bf z}}\left(t\right) + \delta\vec{V}\left(t\right)$, inserting this
expression in the definition of the VAF, and assuming
exponential decay of the orientation, $\langle \hat{{\bf z}}\left(t\right)\cdot
\hat{{\bf z}}\rangle = {\mathrm e}^{-t/\tau_\text{r}}$, and velocity,
$\langle\delta\vec{V}\left(t\right)\cdot\delta\vec{V}\rangle
= (3k_\text{B}T/M_\text{m})\,{\mathrm e}^{-t/\tau_V}$,
correlation functions we obtain,
\begin{equation}
  C_{VV}\left(t\right)
  = \frac{1}{3}\langle V_z\rangle^2 {\mathrm e}^{-t/\tau_\text{r}}
  + \frac{k_\text{B}T}{M_\text{m}} {\mathrm e}^{-t/\tau_\text{v}}.
\end{equation}
Using this expression, the MSD takes the form
\begin{eqnarray}
  \label{eq:MSD-th}
  \Delta L^2\left(t\right)
   &=& 6D_\text{m}t
  - 2\langle V_z\rangle^2\tau_\text{r}^2
  \left(1-{\mathrm e}^{-t/\tau_\text{r}}\right)
  \\
  &&\qquad\qquad
  - 6\frac{k_\text{B}T}{M_\text{m}}\tau_\text{v}^2
  \left(1-{\mathrm e}^{-t/\tau_\text{v}}\right)\nonumber.
\end{eqnarray}
In the ballistic regime, $t\ll\tau_\text{v}$, the MSD reduces to
$\Delta L^2(t) \approx (3k_\text{B}T/M_\text{m}
+ \langle V_z\rangle^2)\,t^2 = {\mathcal B}_\text{I}t^2$, while in the
diffusive regime, $t\gg\tau_\text{r}$, we have $\Delta L^2(t)
\approx 6((k_\text{B}T/M_\text{m})\,\tau_\text{v} +
\frac{1}{3}\langle V_z\rangle^2\tau_\text{r})\,t = 6D_\text{m}t$.
The effective dimer diffusion coefficient is $D_\text{m}=D_0+\frac{1}{3}\langle
V_z\rangle^2\tau_\text{r}$.
The diffusion coefficient in the absence of propulsion is
$D_0 = (k_\text{B}T/M_\text{m})\,\tau_\text{v}$.

\begin{figure}[htbp]
  \input{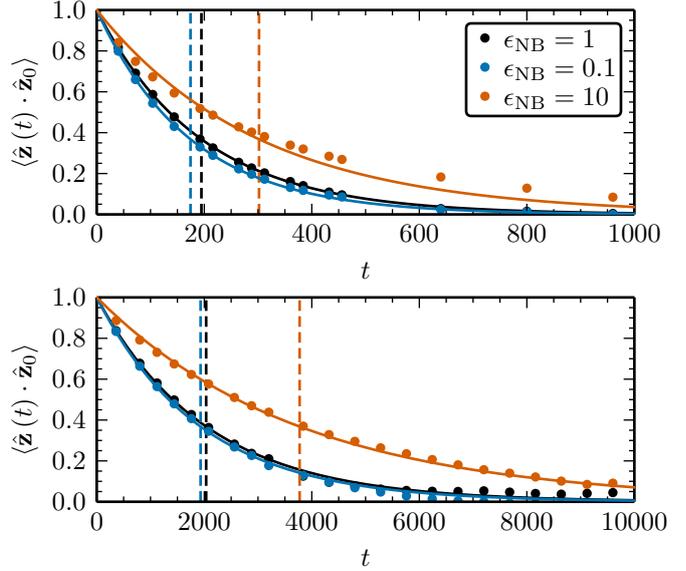}
  \caption{Auto-correlation of orientation $\hat{\mathbf z}\left(t\right)$ for
  small ($\sigma_\text{N} = 4$; top) and large ($\sigma_\text{N} = 10$; bottom)
  dimers, for several values of $\epsilon_\text{NB}$.
  The solid lines show a fit that assumes exponential decay.
  The vertical dashed lines indicate decay times $\tau_\text{r}$.}
  \label{fig:orientation}
\end{figure}

The assumptions made in constructing this model for the VAF and MSD can be
tested by direct simulation.
The orientational correlation function is plotted in fig.~\ref{fig:orientation}
for dimers with $\sigma_\text{N}=10$ and 4 for three values of $\epsilon_\text{NB}$.
The correlation function decay is well approximated by an exponential form and
the orientational relaxation time $\tau_\text{r}$ can be determined.
Note the significant difference in the orientational relaxation times between
the $\epsilon_\text{NB}=0.1$ and $\epsilon_\text{NB}=1$ cases, which are the same within
statistical errors, and the $\epsilon_\text{NB}=10$ case where $\tau_\text{r}$ is
significantly longer.
In this latter case, since $\epsilon_\text{NB} >> \epsilon_\text{NA}$, the B particles
interact with the N sphere with a repulsive potential that is much stronger than
the A--N repulsive potential.
This results in propulsion with the N sphere in front (negative velocity) and
motion down the concentration gradient, which tends to stabilize the motor to
orientational fluctuations.
Such effects have been considered for pushers and pullers for motors with
dimer geometries~\cite{popescu:11}.
The $\tau_\text{r}$ relaxation times for a variety of parameters are presented
in Table~\ref{tab:dimer}.

\begin{figure}[htbp]
  \input{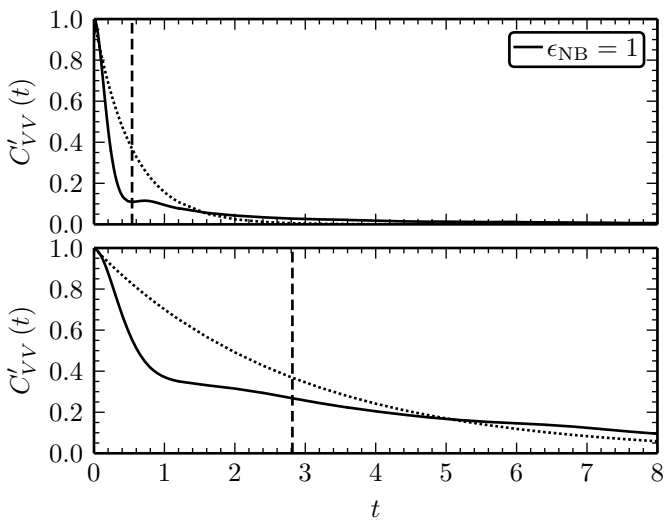}
  \caption{Normalised auto-correlation of the velocity,
  $C_{VV}^\prime\left(t\right) = C_{VV}\left(t\right) / (k_\text{B}T/M_\text{m})$,
  for small ($\sigma_\text{N} = 4$; top) and large ($\sigma_\text{N} = 10$; bottom)
  dimers, for several values of $\epsilon_\text{NB}$.
  The dotted lines show a fit of an exponential decay to the measured data.
  The vertical dashed lines indicate decay times $\tau_\text{v}$.}
  \label{fig:vac}
\end{figure}

The VAFs for an inactive dimer ($\langle V_z \rangle=0$) are presented in
fig.~\ref{fig:vac}.
These correlation functions cannot be quantitatively modelled by exponential
decay and show additional structure.
As expected for systems with continuous potentials there is a very small
inertial regime where the VAF behaves as $\sim \langle V(0)^2 \rangle -\langle
F(0)^2 \rangle/(2M_\text{m}^2) t^2 $, and gives rise to a zero initial slope.
Here $F(0)$ is the initial value of the force on the dimer.
This regime has negligible consequence for our low Reynolds number conditions.
The VAF exhibits other characteristic features.
For the small dimer with $\sigma_\text{N}=4$ a minimum in the decay is clearly
seen and is due to caging effects that are significant when the dimer is
comparable in size to the solvent.
For the larger dimer this effect is smaller and no minimum is seen.
For both dimer sizes there is a long-time power law decay due to the
coupling of solvent collective modes to the dimer centre of mass velocity.
The long-time decay is especially evident for the larger dimer.

The simulation results can now be compared with the theoretical expression in
eq.~(\ref{eq:MSD-th}) and its limits in the ballistic and diffusive regimes.
In the ballistic regime we have ${\mathcal B}'_\text{I}=3+\langle V_z' \rangle^2$.
Table~\ref{tab:dimer} shows that the ballistic regime is dominated by thermal
speed $3k_\text{B}T/M_\text{m}$ (${\mathcal B}'_\text{I} \simeq 3$), rather than
the square of the propulsion velocity as is the case for micron-scale motors.
This is one of the signatures that distinguish the behaviour of
{\AA}ngstr{\"o}m-scale motors from their larger counterparts.
The effects of self-propulsion manifest themselves clearly when the long-time
diffusive regime is examined.
The diffusion coefficient $D_\text{m}$ obtained from the MSD and its theoretical
estimate $D_\text{m}^\text{th}$ are in close accord.
When $\epsilon_\text{NB}=1$ there is no propulsion and the diffusion coefficient
is given by $D_0$.
Propulsion contributes significantly to the magnitude of the diffusion
coefficient and $D_\text{m}$ is 50--100 times larger than $D_0$ for the large
motor and 3--9 times larger for the small motor.
The full expression for $\Delta L^2(t)$ in eq.~(\ref{eq:MSD-th}) agrees well with
the simulation data (fig.~\ref{fig:msd}); the small deviations can be attributed
to the approximation that the VAF decays exponentially.
\pagebreak

Converted to physical units, the large motor with length $(\sigma_\text{C} +
\sigma_\text{N}) / 2 + R = 5.79\,\text{nm}$ is propelled with velocity $\langle
V_z\rangle = 3.67\,\text{nm}\,\text{ns}^{-1}$, for $\epsilon_\text{NB} = 0.1$.
For $\tau_\text{r} = 4.15\,\text{ns}$ it travels an average distance $\langle
V_z\rangle\tau_\text{r} = 15.23\,\text{nm}$, or 2.63 times its length.
The small motor with length $2.55\,\text{nm}$ is propelled with velocity
$4.40\,\text{nm}\,\text{ns}^{-1}$.
For $0.37\,\text{ns}$ it travels an average distance $1.65\,\text{nm}$, or 0.65
times its length.

The results reported here show that {\AA}ngstr{\"o}m-scale chemically powered
motors operate in a strongly fluctuating medium where the ballistic regime is
dominated by the thermal speed rather than the propulsion speed.
Nevertheless, self-propulsion is responsible for the very large enhancements of
the motor diffusion coefficients.
The motors are driven by the microscopic analogues of phoretic mechanisms and,
after substantial averaging to remove the effects of fluctuations, one can
observe the solvent flow fields that are an integral part of the propulsion
mechanism.
From the above discussion we see that these motors operate in nanometre and
nanosecond regimes.
Although the average motor velocities are very large, the reorientation times
are very short.
The results suggest that such very small motors could have applications in
nanoscale systems that make use of the fast motor diffusion and the solvent
velocity fields generated by the motor motion.
If the motion of the motors is constrained so that orientational relaxation is
hindered, their directed motion could be exploited to achieve targeted delivery
or active transport on nanoscale, very much like biological machines.

\acknowledgements
Part of this work was supported by a grant from Natural Sciences and
Engineering Research Council of Canada.

\bibliographystyle{eplbib}
\bibliography{single_dimer}

\end{document}